\documentclass[prd,twocolumn,showpacs,nofootinbib,floats,floatfix]{revtex4}
\usepackage{latexsym}
\usepackage{amsmath,amsfonts}
\usepackage{graphicx} 

\newcommand{\D}{\mathrm{d}}
\newcommand{\EL}{\epsilon_{d}}
\newcommand{\EO}{\epsilon_{\Omega}}
\newcommand{\EE}{\epsilon_{E_{b}}}
\newcommand{\EJ}{\epsilon_{J}}

\begin{document}
\title{Measuring eccentricity in binary black-hole initial data}
\author{Jason D. Grigsby}\email{grigjd3@wfu.edu}
\author{Gregory B. Cook}\email{cookgb@wfu.edu}
\affiliation{Department of Physics, Wake Forest University,
                 Winston-Salem, North Carolina\ \ 27109}
\date{\today}

\begin{abstract}
Initial data for evolving black-hole binaries can be constructed via
many techniques, and can represent a wide range of physical scenarios.
However, because of the way that different schemes parameterize the
physical aspects of a configuration, it is not alway clear what a
given set of initial data actually represents.  This is especially
important for quasiequilibrium data constructed using the conformal
thin-sandwich approach.  Most initial-data studies have focused on
identifying data sets that represent binaries in quasi-circular
orbits.  In this paper, we consider initial-data sets representing
equal-mass black holes binaries in eccentric orbits.  We will show
that effective-potential techniques can be used to calibrate initial
data for black-hole binaries in eccentric orbits.  We will also
examine several different approaches, including post-Newtonian
diagnostics, for measuring the eccentricity of an orbit.  Finally,
we propose the use of the ``Komar-mass difference'' as a useful,
invariant means of parameterizing the eccentricity of relativistic
orbits.
\end{abstract}


\maketitle

\section{Introduction}
The possible detection of gravitational waves by detectors such as
LIGO and LISA is driving rapid progress in the binary black hole (BBH)
problem.  The final stages of the inspiral and coalescence are believed
to be primary sources of gravitational waves at frequencies accessible
by such detectors. Theoretical models that accurately
predict these final stages of inspiral are needed to help analyze and
improve the rate of detection with future data. There are two
techniques commonly used to study such systems, the post Newtonian
approximation and numerical relativity. Since post Newtonian methods
are expected to break down at smaller separations, accurate models of
final stage inspirals need to be handled with numerical techniques.

Numerical relativity breaks the problem into two parts, the specification
of initial data and the numerical time evolutions of this data.
With recent advances in 
evolutions\cite{Pretorius-2005,Campanelli-etal-2006a,Baker-etal-2006a,Diener-etal-2005},
it is as important as ever to fully understand the initial data one is
starting with. Post Newtonian methods have shown that BBHs starting
with large separation will evolve toward an adiabatic 
inspiral that follows a series of quasi-circular orbits\cite{Peters64}. This
quasicircular inspiral is thought to last till late stages where the
black holes plunge toward coalescence.  Most work to date in both the
construction of BBH initial data and evolutions have focused on 
quasi-circular configurations.  But the study of binaries in truly
eccentric orbits near coalescence may be important for gravity-wave
detectors (especially LISA) and, in any case, is of considerable
theoretical interest.  A primary goal of this paper is to take a
preliminary look at how current quasiequilibrium methods for 
constructing BBH initial data can be extended to construct general
eccentric configurations.

To predict quasicircular
orbits, two techniques have commonly been used: the effective-potential 
(EP) method\cite{cook94e} and the Komar-mass ansatz
\cite{gourgoulhon-etal-2002a}.
The Komar-mass ansatz compares two definitions of energy: the ADM
energy\cite{ADM} and the Komar mass\cite{komar59}.  
The ADM energy gives a proper definition of energy at spatial infinity
in all cases, but the Komar mass is only accurate when a configuration 
is at least momentarily stationary.  So, the Komar-mass ansatz
posits that when these two energies are equal, the system should be in
quasiequilibrium.  The EP technique is motivated
by the Newtonian effective one-body problem, and determines a
quasi-circular configuration by finding models that have a
minimum in the binding energy along a sequence of constant angular momentum,
constant masses, and constant spins. 
The two methods have been contrasted and
largely agree, with significant differences occurring only at
separations close to final plunge of the 
BBHs\cite{caudill-etal-2006, skoge-baumgarte-2002}. 

It has been conjectured\cite{caudill-etal-2006} that the EP method,
when used with methods for constructing quasiequilibrium initial data,
should generalize to represent BBHs in eccentric orbits.  For
initial-data methods designed to produce quasiequilibrium data, the
generalization is not obvious because the individual black holes will
not follow the integral curves of any approximate Killing field.  We
will explore this conjecture extensively in this paper.  In order for
this conjecture to be true, the initial data used to construct the
effective potentials must represent BBHs at turning points in their
orbits.  Unfortunately, we cannot test this property of the data
directly.  We will justify the assumption that the data represent BBHs
at turning points and test it by comparison to various results from
post-Newtonian methods.

By constructing effective-potentials that represent BBHs in eccentric
orbits, it becomes possible to make direct estimates of the
eccentricity of an orbital configuration.  Of course, it is not
possible to justify a unique definition for eccentricity for
relativistic configurations where elliptic orbits do not exist.
Instead, we compare several different definitions\cite{mora-will-2004}
and show that they give reasonable and consistent results.  Because
there is no unique definition of eccentricity, it is perhaps not the
best quantity for parameterizing eccentric orbits.  We find that
the ``Komar-mass difference'', the quantity that is set to zero in the
Komar-mass ansatz for defining circular orbits, may serve as a useful
parameterization of eccentric orbits.

We begin in Sec.~\ref{sec:quas-init-data} with an overview of the
quasiequilibrium method we use to construct BBH initial data,
emphasizing the aspects that will be most relevant to our subsequent
discussion.  In Sec.~\ref{sec:effective-potentials}, we will discuss
various aspects of effective potentials in the context of their use
with BBH initial data.  In particular, we will justify as fully as we
can the extension of these effective potentials to BBHs in eccentric
orbits.  In Sec.~\ref{sec:eccentricity} we will explore and compare
several definitions of eccentricity, and will motivate the use of the
``Komar-mass difference'' as an invariant means of parameterizing BBHs
in eccentric orbits.  All of the preceding discussions have dealt
with non-spinning black holes.  In Sec.~\ref{sec:corotation}, we will
briefly revisit the case of corotating black holes.  We end the
paper in Sec.~\ref{sec:conclusions} with some conclusions.

\section{Initial Data}
\label{sec:quas-init-data}

The binary black-hole initial-data sets that are used below were
described in detail in
Refs.~\cite{caudill-etal-2006,cook-pfeiffer-2004a} and references
within.  Here, we give an overview of the methods used to construct
the initial data, elaborating only on the details most relevant to
finding black-hole binaries in quasi-circular orbits.

Our initial-data sets are constructed within the {\em extended
conformal thin-sandwich} approach\cite{york-1999,Pfeiffer-York-2003}.
This approach is based on the standard $3+1$ decomposition where the
space-time interval is written as
\begin{equation}
\D{s}^2 = -\alpha^2\D{t}^2+\gamma_{ij}(\D{x}^i+\beta^i\D{t})
                                      (\D{x}^j+\beta^j\D{t}).
\end{equation}
Here, $\gamma_{ij}$ is the spatial metric, and $\alpha$ and
$\beta^{i}$ are the lapse function and shift vector.  Minimal initial data
for a Cauchy evolution requires that we fully specify $\gamma_{ij}$
and the extrinsic curvature $K_{ij}$ (essentially a first time
derivative of the spatial metric) defined as
\begin{equation}
K_{ij} \equiv -\frac{1}{2}{\cal L}_{n}\gamma_{ij},
\end{equation}
where $n^\mu$ is the time-like unit normal to the $t=\mbox{const.}$
initial-data surface.

The conformal thin-sandwich (CTS) approach requires a conformal
decomposition of $\gamma_{ij}$, and that we specify the conformally
related metric $\tilde\gamma_{ij}$.  In this work we will always take
$\tilde\gamma_{ij}$ to be flat.  We must also specify the time
derivative of the conformal metric $\partial_t\tilde\gamma_{ij}$ and
the trace of the extrinsic curvature.  We fix both quantities to be
zero.  The CTS approach then requires that we determine the conformal
factor $\psi$ relating $\gamma_{ij}$ and the conformal metric, and the
shift vector $\beta^i$.  These are obtained by solving elliptic
versions of the Hamiltonian and momentum constraint equations (see
Ref.~\cite{caudill-etal-2006} for details). Finally, the extended CTS
approach also requires that we determine the lapse function $\alpha$
by fixing the time derivative of the trace of the extrinsic curvature
and then solving the evolution equation for the trace of the extrinsic
curvature as an elliptic equation.  We fix the trace of the extrinsic
curvature to be constant in time.

In constructing black-hole initial data, we excise the black-hole
interior from the computational domain and must impose boundary
conditions at the excision surfaces when solving the elliptic
equations for $\psi$, $\beta^i$, and $\alpha$.  We demand that each
black hole be in quasiequilibrium by imposing the boundary conditions
worked out in Refs.~\cite{Cook-2002} and \cite{cook-pfeiffer-2004a}.
The assumptions of quasiequilibrium are essentially the same as those
required of an ``isolated horizon'' (see
Refs.~\cite{ashtekar-krishnan-2003a,ashtekar-etal-2000a,Jaramillo-etal:2004}).
We must also impose boundary conditions at the outer boundary of the
computational domain (either at infinity or some large radial distance
from the black holes).  For this, we assume that our configuration
is asymptotically flat.  However, asymptotic flatness does not fully
fix the boundary conditions on the constrained data.  The asymptotic
condition on the shift is
\begin{eqnarray} \label{eq:shiftBC}
\beta^i|_{r\rightarrow\infty} = (\mathbf{\Omega}_0 \times \mathbf{r})^i,
\end{eqnarray}
where $\mathbf{\Omega}_0$ is an angular velocity vector.

The time coordinate threading through our initial-data slice is defined
by
\begin{equation}
t^\mu \equiv \alpha n^\mu + \beta^\mu.
\end{equation}
Imposing Eq.~(\ref{eq:shiftBC}), we see that $\mathbf{\Omega}_0$
determines the rotation of the ``helical'' time coordinate.  For a binary
system in equilibrium, the time coordinate generates a symmetry and the
bodies move in circular orbits along integral lines of the time
coordinate.  For relativistic systems, the binary can only be in
quasiequilibrium and the time coordinate generates an approximate
symmetry.  Nevertheless, $\mathbf{\Omega}_0$ represents the orbital
angular velocity of the binary as measured by observers at infinity.
From a computational perspective, $\mathbf{\Omega}_0$ must be chosen.
Ultimately, it is the effect of different choices for $\mathbf{\Omega}_0$
that we will be exploring in this paper.

In Ref.~\cite{caudill-etal-2006}, two independent methods for choosing
the magnitude of $\mathbf{\Omega}_0$ were compared.  Both methods
attempt to produce a binary system that is in quasiequilibrium with
the black holes in quasi-circular orbits.  One method is based on the
Komar-mass ansatz, first proposed by Gourgoulhon et.\
al.\cite{gourgoulhon-etal-2002a}, which posits that if $\Omega_0$ is
chosen so that the ADM energy $E_{\mbox{\tiny ADM}}$ and the Komar mass
$M_{\mbox{\tiny K}}$ of a system are equal, then the system will be
nearly stationary (i.e.\ in quasiequilibrium) and the binary will be
in a quasi-circular orbit.  These masses are defined via
\begin{eqnarray}
\label{eqn:ADM_mass}
E_{\mbox{\tiny ADM}} &=& \frac{1}{16\pi}\oint_{\infty}\nabla_j\left({\cal G}^j_i-\delta^j_i {\cal G}\right)d^2S^i, \\
\label{eqn:Komar_mass}
M_{\mbox{\tiny K}} &=& \frac{1}{4\pi}\oint_{\infty}\left(\nabla_i\alpha - \beta^jK_{ij}\right)d^2S^i,
\end{eqnarray}
where ${\cal G}_{ij} \equiv \gamma_{ij} - f_{ij}$, $f_{ij}$ is the
flat metric, and $\nabla_j$ is the covariant derivative compatible
with $\gamma_{ij}$.  The second method assumes that quasi-circular orbits
are found at the minima of a reduced two-body effective potential.
Effective potentials will be described in more detail in
Sec.~\ref{sec:effective-potentials}.

\begin{figure}[ht]
\includegraphics[width=\linewidth,clip]{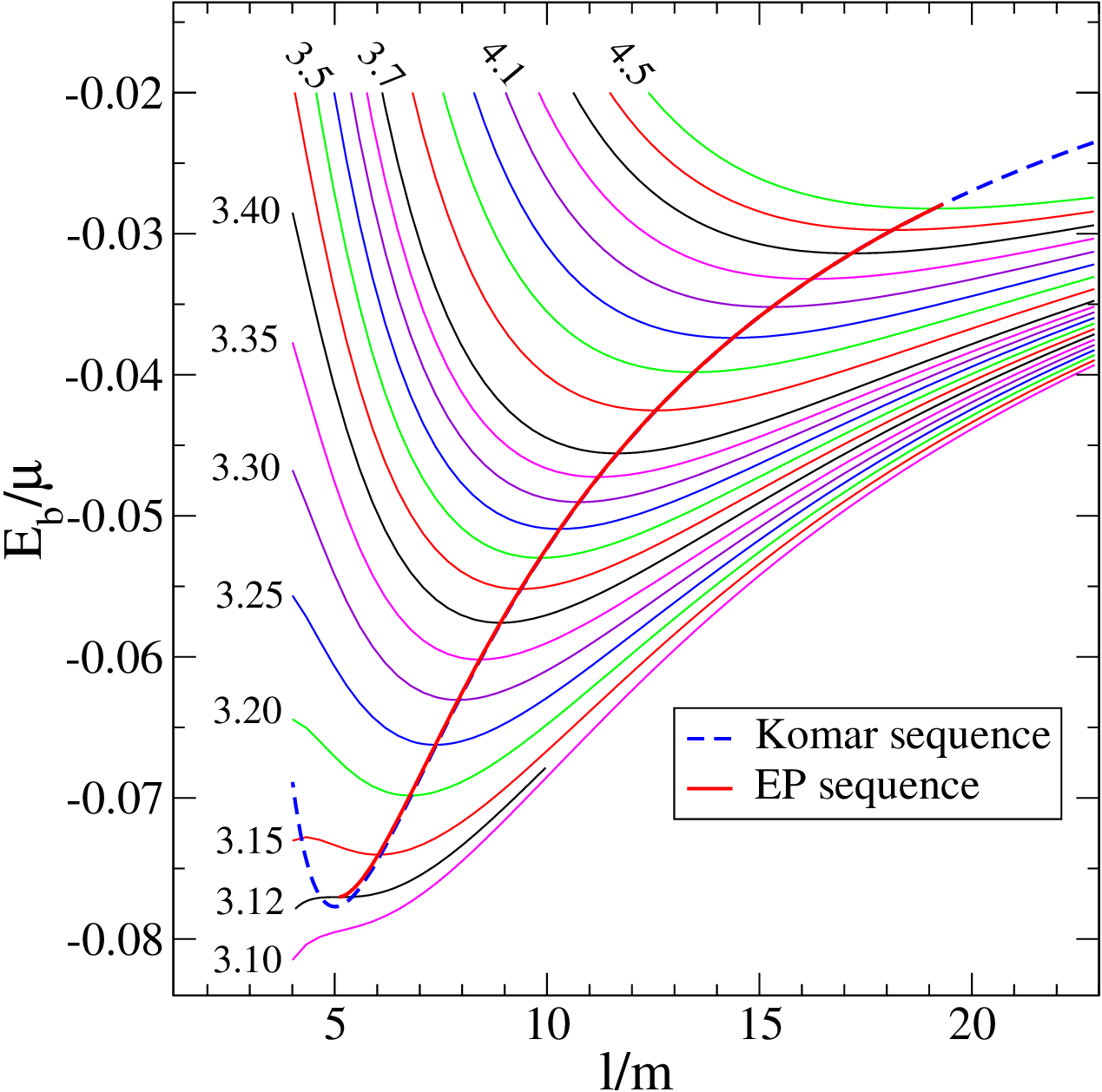}
\caption{\label{fig:num_non} Effective-potential plot for non-spinning
equal-mass black holes constructed from numerical initial
data\cite{caudill-etal-2006}.  The thin solid(multi-color) lines, are
individual EP curves.  Some of these EP curves are labeled it value of
$J/\mu{m}$.  Passing through the local minima of these EP curves and
drawn as a bold(red) line is the EP sequence of quasi-circular
orbits.  The Komar sequence of quasi-circular orbits is displayed as a
dashed(blue) line.}
\end{figure}
In Ref.~\cite{caudill-etal-2006}, we showed that the circular-orbit
configurations produced by both methods agree remarkably well for both
non-spinning and corotating black-hole binaries. Here, we simply show
a figure that directly compares sets of circular-orbit models as we
vary the binary separation.  Figure~\ref{fig:num_non} displays the
effective potential (EP) curves for non-spinning equal-mass black-hole
binaries as computed in Ref.~\cite{caudill-etal-2006}.  The vertical
axes display the dimensionless binding energy of the binary $E_{\rm
b}/\mu$, where $E_{\rm b}$ is defined in Eq.~(\ref{eq:E_b}),
$\mu\equiv m_1m_2/m$ is the reduced mass of the system and $m\equiv m1+m2$
is the total mass of the binary.  The
horizontal axes display the dimensionless proper separation of the
binary $\ell/m$.  It is
important to note that each EP curve consists of a sequence of models
where the value of $\Omega_0$ changes monotonically.  Passing through
the minima of the EP curves is the ``EP sequence'' defined as the
sequence of quasi-circular orbit models where quasi-circular orbits
are defined via the effective-potential method.  Also shown in
Fig.~\ref{fig:num_non} is the ``Komar sequence'' defined as the sequence of
quasi-circular orbit models defined via the Komar-mass ansatz.  It is
clear that the two sequences agree quite well except for the smallest
separations.  See Ref.~\cite{caudill-etal-2006} for a more detailed
comparison.  Most important for our considerations is that both the
Komar-mass and EP methods choose particular models on each EP curve,
with very similar values of $\Omega_0$, as quasi-circular orbit
models.

These sequences of quasi-circular orbit models represent a good
approximation to the adiabatic inspiral of a black-hole binary so long
as the binary separation is not too small.  For large enough
separation, the time scale for radiation reaction to induce a
significant change in the orbital radius is much larger than the
orbital period.  At small enough separations, radiation reaction
becomes significant and the sequences of quasi-circular obits will
eventually become a poor approximation to an adiabatic inspiral.  It
is difficult to estimate exactly where this transition occurs, but it
will certainly occur at separations larger than the inner-most stable
circular orbit (ISCO).  For a Komar sequence, the ISCO is defined to
occur at the minimum in the binding energy.  For an EP sequence, the
ISCO is defined to occur at the point where the EP curves no longer
have a local minimum.  Both ISCOs occur near $\ell/m\sim5$ (see
Fig.~\ref{fig:num_non}).

When the CTS equations and excision boundary conditions as described
above are used to construct initial data for binaries in
quasi-circular orbit, the resulting data are consistent with a system
in quasiequilibrium.  In particular, we have both bodies in the binary
following the integral curves of an approximate helical Killing vector
(the time coordinate).  Furthermore, half of the initial data that
must be specified is also consistent with this notion of
quasiequilibrium.  Recall that the conformal metric
$\tilde\gamma_{ij}$, the trace of the extrinsic curvature, and their
time derivatives must be specified, and that {\em we take all of the time
derivatives to vanish}.  As long as the binary is in a circular orbit,
the notion of quasiequilibrium is satisfied.  

However, we are left to wonder, what happens if we choose $\Omega_0$
so that the binary is {\em not} in a quasi-circular orbit?  In
Ref.~\cite{caudill-etal-2006}, we made the assertion that the
resulting initial data would represent a binary at either pericenter
or apocenter of a general bound or unbound orbit.  To arrive at this
conclusion, we are forced to give up the notion that the helical time
vector represents an approximate Killing vector of the space-time.
Our goal below is to see if this interpretation is reasonable.

\section{Effective Potentials}\label{sec:effective-potentials}

In Newtonian physics, 1-D effective potentials nicely capture the
important features of certain dynamical systems. In the case of the
reduced gravitational 2-body problem, for a given orbital angular
momentum, the effective potential can be used to locate the turning
points for an elliptic orbit of given energy, or the radius and energy
of a circular orbit.  No such 1-D effective potential can be
rigorously and uniquely derived for the fully relativistic
gravitational 2-body problem.  However, a useful effective potential
has been defined\cite{cook94e, baumgarte-2000, pfeiffer-etal-2000,
skoge-baumgarte-2002} in direct analogy with the Newtonian
gravitational effective potential.

In essence, an EP curve is the total energy of the system for a
sequence of configurations where the radial separation varies, but all
other physical parameters are held fixed.  In order to correspond to
an effective potential, the velocity of the generalized 
coordinate that is allowed
to vary (the radial separation in this case) must vanish so that there
is no associated kinetic energy.  Binary systems in bound orbits with
vanishing radial velocity are either at apocenter or pericenter, and
collectively we refer to these as turning points in the orbit.

For black-hole binaries, we use the following functional definition
for an EP curve.  We consider a sequence of initial-data
configurations that hold constant the apparent-horizon masses of the
individual black holes ($m_1$ and $m_2$), 
the magnitude and direction of the spins of
the individual black holes, and the total angular momentum of the
system\footnote{It would be preferable to hold the total mass
of the individual black holes constant.  However, as we will discuss
 in the main text, this is not possible.}.  
We also require that the binary be at a turning point in the
orbit.  For the value of the effective potential, we could simply use
the total ADM energy of the system.  However, it is more intuitive to
use the binding energy since this allows us to estimate if a given
configuration is in a bound orbit.  The binding energy requires that
we subtract from the total energy, the total mass of the system at
infinite separation (i.e.\ the sum of the masses of the individual
black holes).  Unfortunately, there is no rigorous way of defining the
total mass of an individual black hole at finite separation.  This is
a fundamental uncertainty in defining a relativistic effective
potential.

The total mass of an individual black hole includes both its
irreducible mass and the kinetic energy associated with the spin.  In
general, we cannot rigorously define the spin of an individual black
hole, nor do we know how to compute the total mass of an individual
black hole given the spin and apparent-horizon mass.  A reasonable
approximation is to define the total mass of an individual black
hole, $M$, via the Christodoulou formula\cite{christ70}  
\begin{equation}\label{eq:Mtot}
M^2 = M^2_{\mbox{\tiny irr}} + \frac{S^2}{4M^2_{\mbox{\tiny irr}}}
\end{equation}
where $S$ is the magnitude of the spin of the black hole and 
we can approximate the irreducible mass, $M_{\mbox{\tiny irr}}$,
with the apparent-horizon mass.

To avoid some of these ambiguities, we will define the binding energy
as the ADM energy minus the sum of the apparent horizon masses.
\begin{equation}\label{eq:E_b}
E_b \equiv E_{\mbox{\tiny ADM}}-m_1-m_2
\end{equation}
Because
we keep the apparent-horizon masses constant along an EP curve,
this constant shift does not affect the effective potential.  But,
since we do not keep the total mass constant along an EP curve, the
uncertainty in the definition of the total mass contributes to the
uncertainty in the definition of an EP curve.

Another fundamental uncertainty in our numerical EP curves is the assumption
that they are constructed of turning points in eccentric orbits.  When
the EP method was introduced, it was applied while using the conformal
imaging decomposition which allowed one to directly set the radial
momentum of the two holes to zero\cite{cook94e}.  
In the CTS decomposition, there is
no mechanism to directly set such data.  However, as discussed at the
end of Sec.~\ref{sec:quas-init-data}, we conjecture that our initial
data does have zero radial momentum.  The reasoning behind this
conjecture is that the time derivative of the conformal metric is
still taken to vanish, so that at least instantaneously we expect
each black hole to be at a turning point.

\begin{figure}[ht]
\includegraphics[width=\linewidth,clip]{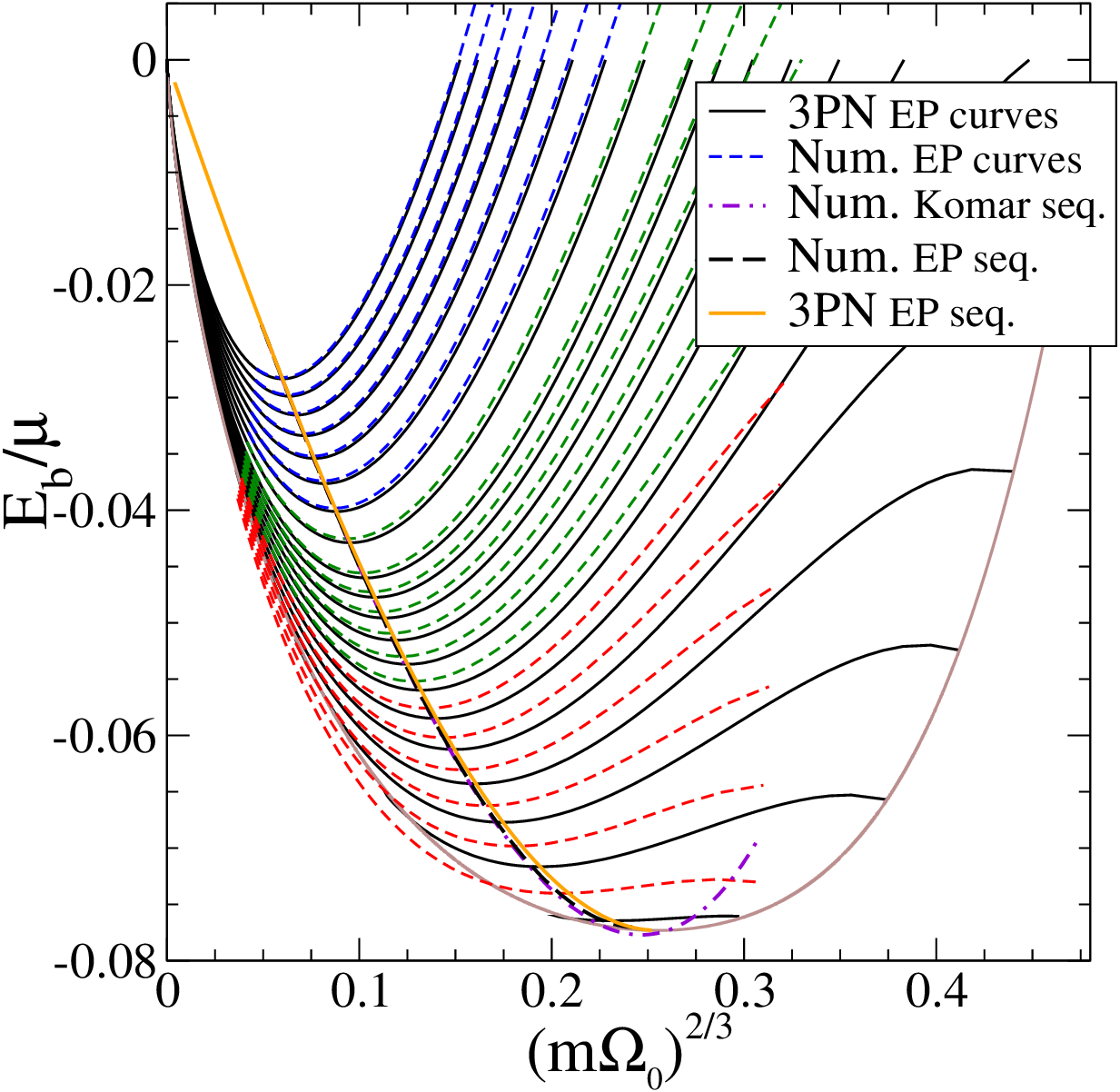}
\caption{\label{fig:Non_const_J} EP curves for non-spinning equal-mass
black holes from both numerical and 3PN data.  Numerical EP curves are
plotted as short-dashed(blue, green, and red) lines.  3PN EP curves
are plotted as solid(black) lines.  The Komar sequence through the
numerical data is plotted as a dash-dot(purple) line.  The EP sequence
through the numerical data is plotted as a long-dash(black) line.  The
EP sequence through the 3PN data is plotted as a light-solid(orange)
line.  A boundary of the allowable region for the 3PN equations is
shown as a light-solid(brown) line.}
\end{figure}

In order to test the validity of this conjecture, we compare our
numerical effective potentials to post Newtonian data.  Mora and
Will~\cite{mora-will-2004} introduced third order conservative post
Newtonian equations for the energy and angular momentum of a system in
terms of an eccentricity $\epsilon$ and inverse semilatus rectum
$\zeta$.  The authors treated the black holes as having zero spin and
ignored dissipative terms.  One can find the orbital angular velocity
at either pericenter or apocenter through the author's choice of
definition of $\epsilon$ and $\zeta$ used in creating these equations.
In order to include the spin of the black holes, it becomes necessary
to include correction terms that create equations of the following
form.
\begin{eqnarray}\label{eq:total_E}
E(\epsilon,\zeta,\omega) &=& E_{\mbox{\tiny ADM}}(\epsilon,\zeta) 
+ E_{\mbox{\tiny self}}(\epsilon,\zeta,\omega) \\ \nonumber &&\mbox{}
+ E_{\mbox{\tiny N,Corr}}(\epsilon,\zeta,\omega) 
+ E_{\mbox{\tiny Spin}}(\epsilon,\zeta,\omega) \\
\label{eq:total_J}
J(\epsilon,\zeta,\omega) &=& J_{\mbox{\tiny ADM}}(\epsilon,\zeta) 
+ S(\epsilon,\zeta,\omega) \\ \nonumber &&\mbox{}
+ J_{\mbox{\tiny N,Corr}}(\epsilon,\zeta,\omega) 
+ J_{\mbox{\tiny Spin}}(\epsilon,\zeta,\omega)
\end{eqnarray}
In these equations $\omega$ represents the spin angular velocities of
the black holes\footnote{We use a single parameter $\omega$ for
simplicity.  In general, specifying the spin of two black holes would
require six parameters.}.  The self energy and spin terms 
($E_{\mbox{\tiny self}}$ and $S$) are derived as
expansions of the Kerr formulae relating mass, spin, and rotational angular
velocity. The Newtonian
correction terms ($E_{\mbox{\tiny N,Corr}}$ and $J_{\mbox{\tiny N,Corr}}$) 
stem from the conversion of total mass to irreducible
mass and the ``Spin'' terms 
($E_{\mbox{\tiny Spin}}$ and $J_{\mbox{\tiny Spin}}$) 
represent spin-orbit effects. The parameter
space of these equations are shown below in
Figs.~\ref{fig:3PN_lines_J} and \ref{fig:3PN_lines}.

With this parametrization, one can easily construct 3PN EP curves
using straight-forward one-dimensional root finding.  To create a
sequence of constant angular momentum, one can apply a sequence
of values for the eccentricity to Eq.~(\ref{eq:total_J}), and for
each find the corresponding $\zeta$ that returns the
desired value of angular momentum.  One can then use those values of 
$\epsilon$ and $\zeta$ in Eq.~(\ref{eq:total_E}) to find the energy of 
the system. In
Fig.~\ref{fig:Non_const_J}, we plot for comparison the EP curves
for non-spinning, equal-mass binaries from both the numerical data and
the 3PN equations.  The energy is plotted as a function of the
dimensionless orbital angular velocity $m\Omega$, where small
$m\Omega$ corresponds to large orbital separation.  Included on the
graph are data from the Komar sequence and the EP sequences extracted
from the minima of the 3PN EP curves and the numerical EP curves.  

At large values of angular momentum (large orbital separation), the
solid 3PN data agrees well with the dashed numerical data.  The
numerical data and the 3PN data begin to diverge as the angular
momentum decreases.  This is not surprising as it is well known the PN
expansion is less accurate for tighter binary systems and the same is
true for the numerical initial data models.  It seems from the good
agreement between numerical and 3PN EP curves that the numerical data
we construct using the CTS approach are reasonably close to turning
points, and seem to asymptote to turning points as the system becomes
more Newtonian.

\section{Measuring Eccentricity}\label{sec:eccentricity}

The comparison of our numerical EP curves with 3PN EP curves
in Sec.~\ref{sec:effective-potentials} gives us confidence that 
the numerical data represent systems at either pericenter or apocenter.  
Since each numerical EP curve is constructed from a
sequence of models where the mass, angular momentum and
spins are held constant and are at turning points, it is reasonable to
use these curves to try to define the eccentricity for a given
model. For Newtonian binaries, eccentricity can be measured using
relative separation at pericenter $d_p$ and apocenter $d_a$:
\begin{equation}\label{eqn:eld}
\EL \equiv \frac{d_{a} - d_{p}}{d_{a} + d_{p}}.
\end{equation}
We use $\EL$ to denote the eccentricity of numerical models based
purely on separations.

By applying Newtonian equations of motion, one can replace the above
definition of eccentricity dependent on relative separation with 
a version that is dependent on the
orbital angular velocities at pericenter $\Omega_p$ and apocenter
$\Omega_a$.  Not only is the eccentricity found in those terms, but
the dimensionless inverse semilatus rectum can also be found using the
same parameters\cite{mora-will-2004}.
\begin{eqnarray}\label{eqn:E_O}
\EO &\equiv& \frac{\Omega_{p} - \Omega_{a}}{\Omega_{p} + \Omega_{a}}, \\
\label{eqn:zeta}
\zeta &\equiv& \left(\frac{\sqrt{m\Omega_p}+\sqrt{m\Omega_a}}{2}\right)^{4/3}.
\end{eqnarray}
We use $\EO$ to denote the eccentricity of our numerical models as
measured using orbital angular velocities.  The following relationships
follow directly from these equations:
\begin{equation}\label{eq:zet_ep_Om}
\zeta = \left(\frac{m\Omega_{p}}{(1+\epsilon)^{2}}\right)^{2/3} = \left(\frac{m\Omega_{a}}{(1-\epsilon)^{2}}\right)^{2/3}.
\end{equation}
We note that these equations (at either pericenter or apocenter) are the
only places in any of our work where the sign of the eccentricity
matters and so for the remainder of the paper we will ignore the middle
(pericenter) relationship and {\em assume eccentricity is
negative at pericenter}.  This has the added benefit of cleaning up
many of the figures below.

\subsection{Direct Measurement}

\begin{figure}[th]
\includegraphics[width=\linewidth,clip]{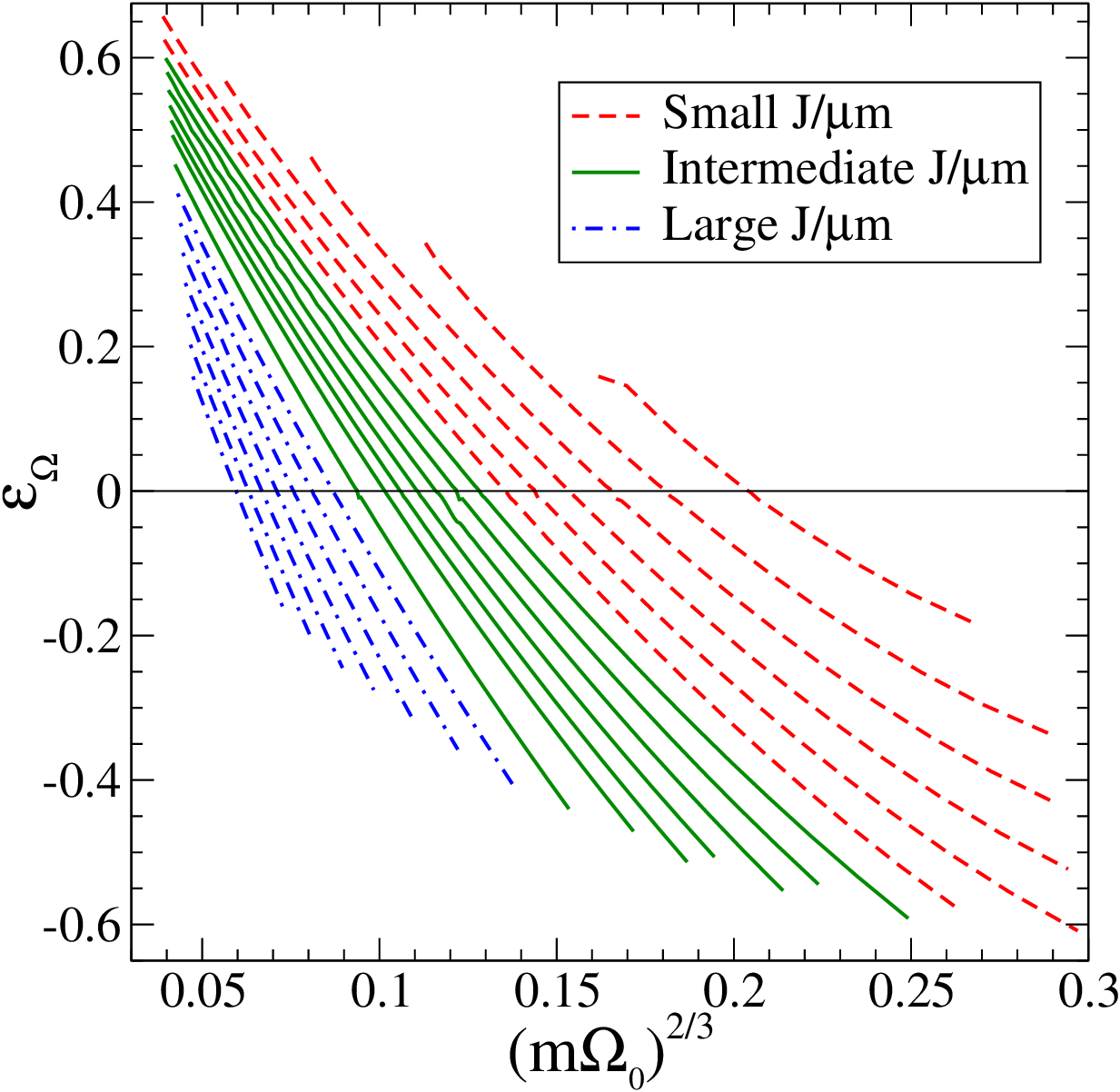}
\caption{\label{fig:non_newt} Eccentricity of non-spinning equal-mass
black holes computed along numerical EP curves.  The $\EO$ definition
of eccentricity is plotted against the orbital angular velocity.  Negative
values of $\EO$ correspond to models at pericenter, while positive
values correspond to apocenter.  Large values of the orbital angular
momentum are plotted as dot-dashed(blue) lines, intermediate values
as solid(green) lines, and small values as dashed(red) lines.}
\end{figure}

Our next goal is to determine the eccentricities of the data in our EP
curves and determine if the definition of eccentricity in
Eq.~(\ref{eqn:E_O}) yields reasonable results.  We start by computing
$\EO$ for the non-spinning equal-mass models represented in the
numerical EP curves displayed in Fig.~\ref{fig:Non_const_J} (also seen
in Fig.~\ref{fig:num_non}).  The measured eccentricities are shown in
Fig.~\ref{fig:non_newt}, plotted against $m\Omega$. The $\EO$
definition of eccentricity should give reasonable results when the
corresponding orbit is sufficiently Newtonian.  The orbits become more
relativistic as the total angular momentum associated with the orbit
gets smaller. Throughout this paper, we will use the following
convention to easily differentiate which EP curves represent large,
intermediate and small values of angular momentum.  EP curves with
dot-dashed lines represent large values of angular momentum, solid
lines denote the middle range of angular momentum, and the dashed
lines have small values of angular momentum.

Figure~\ref{fig:non_newt} shows only a limited range of eccentricities
for each value of angular momentum.  There are several reasons for
this, both physical and computational.  First, to compute the
eccentricity, we require data from two corresponding turning points on
the same EP curve.  That is, we need two points with the same value of
the binding energy.  Because we do not construct models at arbitrarily
large separation, some data at pericenter have no matching data at
apocenter.  In this case, the eccentricity cannot be computed.
Clearly, we cannot compute eccentricities using this method for
pericenter data corresponding to unbound orbits, but there there are
additional limitations associated with the shape of the effective
potential at small separation.  Because of strong-field effects, the
effective potentials reach a local maximum at small separation (cf.\
the effective potential for massive test particles orbiting
Schwarzschild).  For bound orbits, we are limited to computing
eccentricities for configurations with energies lower than that of the
local maximum for each EP curve.  So, the eccentricities plotted in
Fig.~\ref{fig:non_newt} correspond to data in the neighborhood of the
local minimum of each EP curve and extending only as far as the lowest
local maximum on either side.

\begin{figure}[th]
\includegraphics[width=\linewidth,clip]{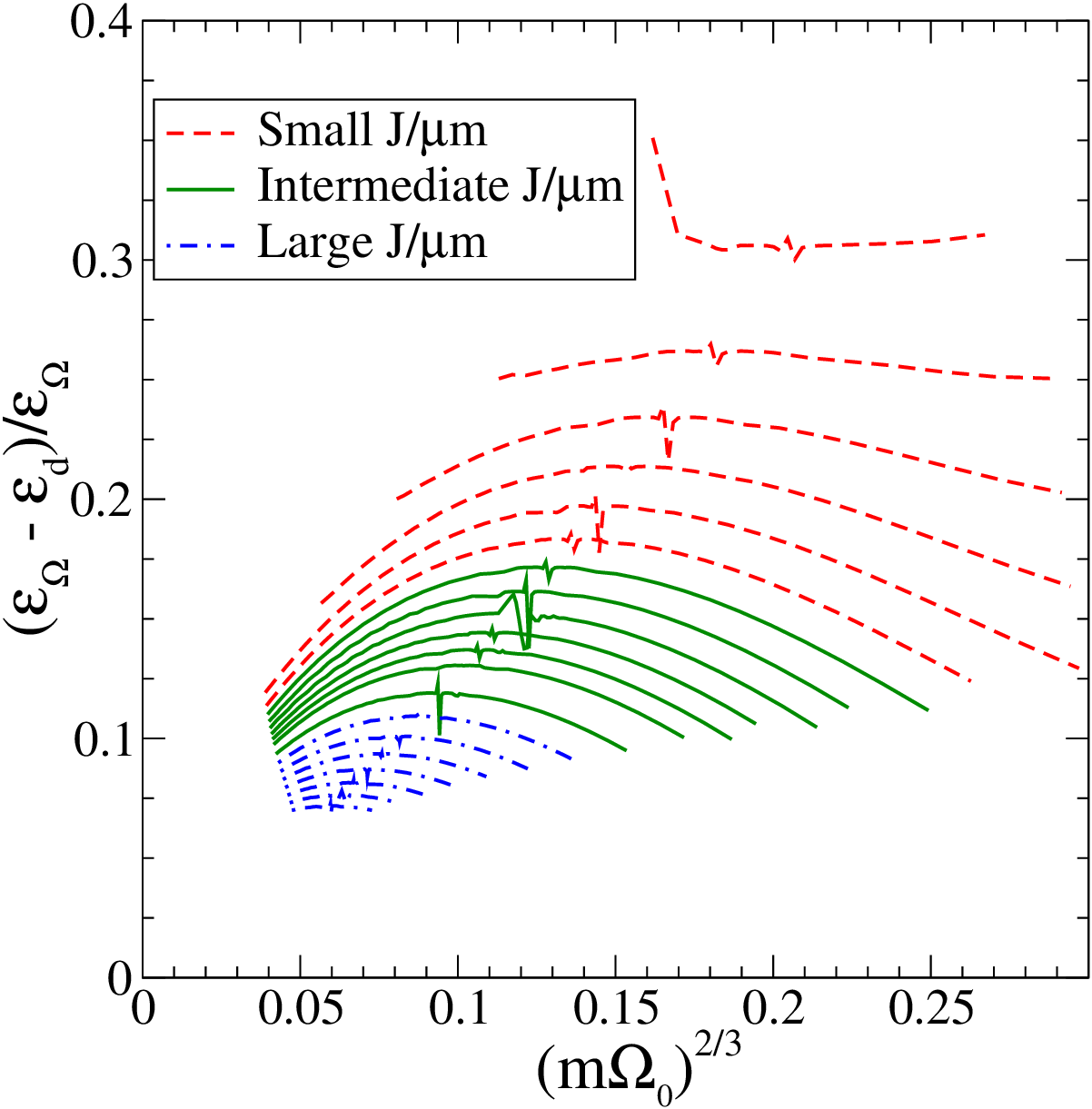}
\caption{\label{fig:non_diff_geom} The relative difference between the
$\EO$ and $\EL$ definitions of eccentricity for the non-spinning
equal-mass black hole numerical EP curves.  Lines as in
Fig.~\ref{fig:non_newt}.}
\end{figure}

To test whether the numerical data is behaving as expected, we turn to
the definition of eccentricity found in $\EL$. This definition should
be reasonable for large separations, but will break down as the coordinate 
separation of the two black holes decreases because of coordinate 
effects near the black holes. In Fig.~\ref{fig:non_diff_geom}, we show the
relative difference of the two measurements $(\EO-\EL)/\EO$.  Note
that we have manually removed data corresponding to points near the
minima of the EP curves since both definitions of eccentricity yield
zero at the minimum and the relative error for neighboring points is
dominated by numerical noise.  However, it is easy to find where those
points would have been as the different lines become somewhat jagged
in the region of zero eccentricity.

Figure~\ref{fig:non_diff_geom} shows the behavior we would have
expected.  For large separations (large $J/\mu{m}$) the coordinate
separation $d$ and the orbital angular velocity should both yield
reasonable estimates of the eccentricity and we see that the relative
error is tending to zero as $J/\mu{m}$ increases.  Clearly, the gauge
dependence of the coordinate separation $d$ will cause $\EL$ to become
less reliable for smaller separations(small $J/\mu{m}$) and indeed, we
see the relative difference increase as $J/\mu{m}$ decreases.  Because
the orbital angular velocity $\Omega$ is gauge independent, we expect
that $\EO$ will yield a better definition of the eccentricity,
however, we need an independent standard against which we can measure
the reliability of $\EO$.

\subsection{Post-Newtonian Measurement}

\begin{figure}[th]
\includegraphics[width=\linewidth,clip]{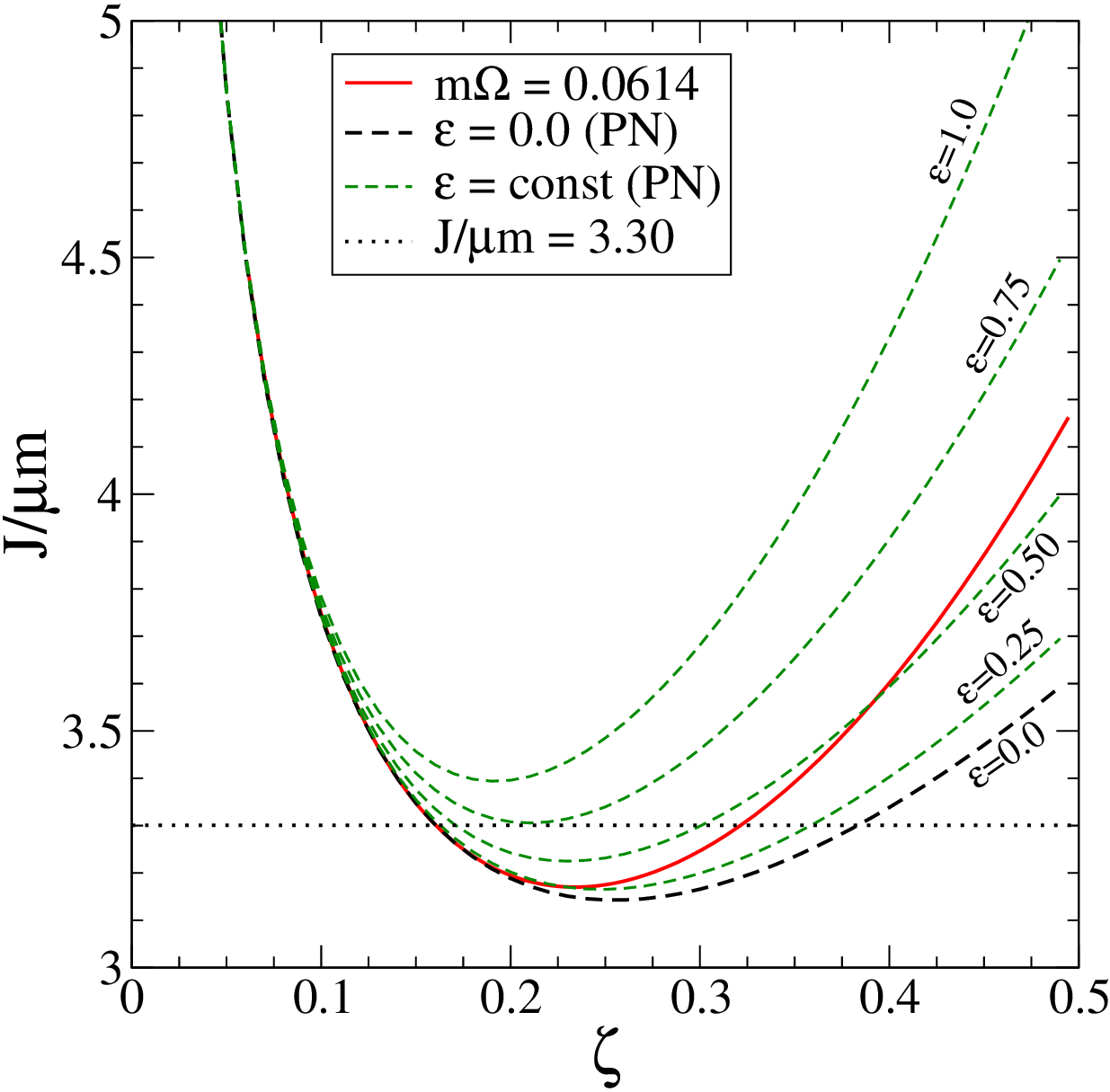}
\caption{\label{fig:3PN_lines_J} Parameter space of
$J(\epsilon,\zeta)$ (Eq.~(\ref{eq:total_J})) for non-spinning
equal-mass black holes.  Short-dashed(green or black) lines of
constant $\epsilon$ are constructed using the 3PN equations.  The
solid(red) line represents all 3PN models with a specific value of
$m\Omega=0.0614$ corresponding to a particular numerical model.  The
horizontal dotted(black) line represents the corresponding orbital
angular momentum of that model.}
\end{figure}
\begin{figure}[th]
\includegraphics[width=\linewidth,clip]{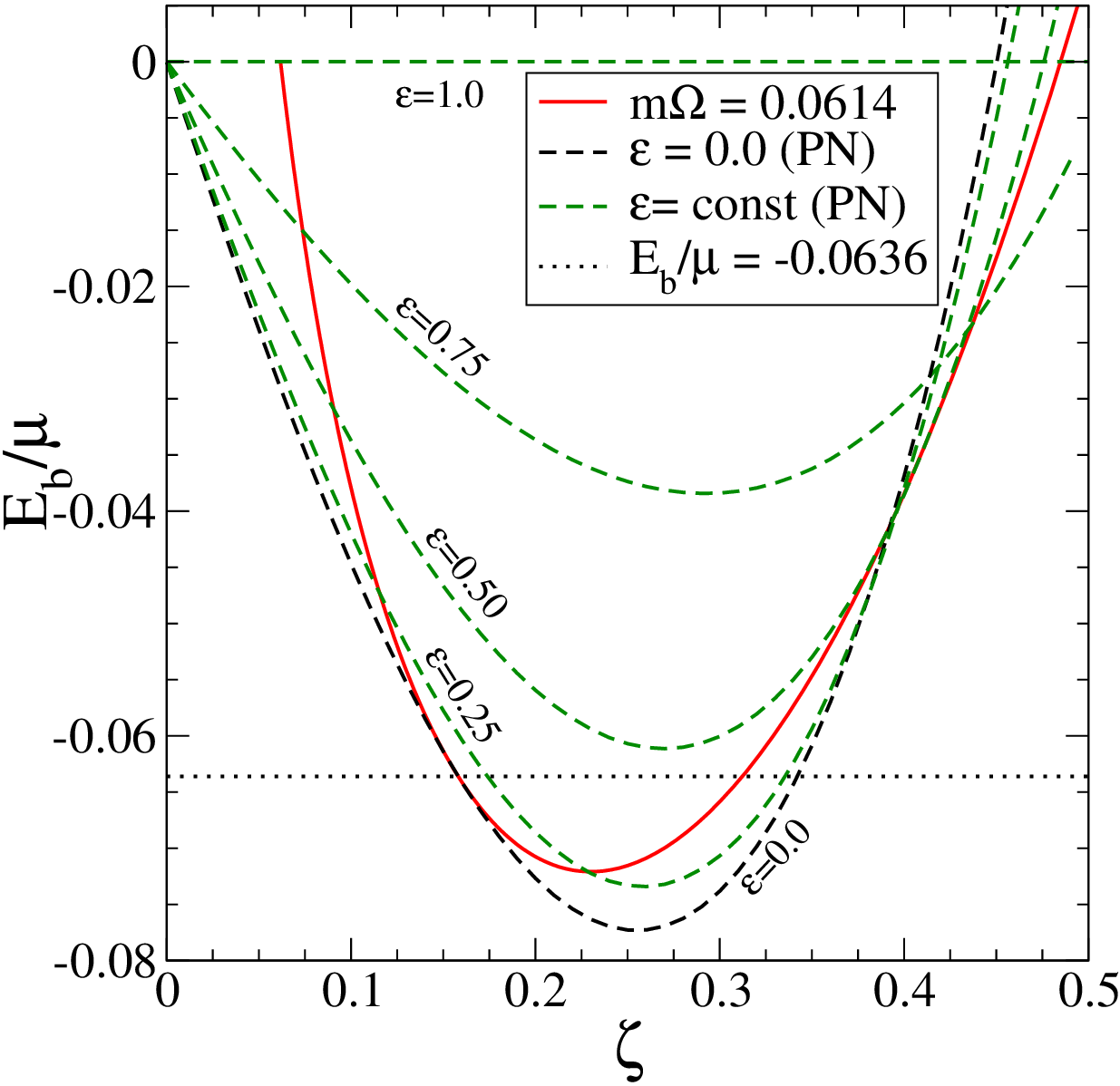}
\caption{\label{fig:3PN_lines} Parameter space of $E(\epsilon,\zeta)$
(Eq.~(\ref{eq:total_E})) for non-spinning equal-mass black holes.
Lines as in Fig.~\ref{fig:3PN_lines_J} except the horizontal
dotted(black) line represents the corresponding binding energy
of the numerical model.}
\end{figure}

To test $\EO$ in the more relativistic regime we return to
post-Newtonian theory.  The 3PN equations for the energy and angular
momentum in Eqs.~(\ref{eq:total_E}) and (\ref{eq:total_J}) can be used
to compute the eccentricity of initial data in several ways, all using
information from a single initial-data
configuration\cite{mora-will-2002,mora-will-2004,BertiIyerWill-2006a}.
The two most useful ways are based on using values for either $E_b$
and $\Omega$, or $J$ and $\Omega$ from a given initial-data set.  Both
methods yield similar but distinct values for the eccentricity.  We
will denote eccentricities obtained using the energy via $\EE$, and
using the angular momentum via $\EJ$.

To use Eqs.~(\ref{eq:total_E}) and (\ref{eq:total_J}) to find
eccentricity, we need to simplify the dependencies.  Those equations
depend on the eccentricity $\epsilon$, the inverse semilatus rectum
$\zeta$, and the individual black-hole spins (represented by
$\omega$).  The spin dependence can always be fixed.  For now, we
consider non-spinning black holes.  Next we apply
Eq.~(\ref{eq:zet_ep_Om}) to replace $\epsilon$ with $m\Omega$ and
$\zeta$.  The equations now depend on only $m\Omega$ and $\zeta$. To
find a 3PN value of $\zeta$, we set one of the equations (say the
energy equation) to the constant (energy) taken from an initial-data
set and replace $m\Omega$ with its value from the same data set.  We
can then use one-dimensional root finding to obtain a value for
$\zeta$.  Finally, using Eq.~(\ref{eq:zet_ep_Om}) again, we can obtain
$\epsilon$.

There is an issue when using root-finding methods on the modified
equations.  Eqs.~(\ref{eq:total_E}) and (\ref{eq:total_J}) are
polynomials with multiple roots, so we must determine which value of
$\zeta$ to use.  Figure~\ref{fig:3PN_lines_J} shows the parameter
space for the 3PN angular momentum from Eq.~(\ref{eq:total_J}) for the
case of non-spinning equal-mass black holes.  The angular momenta for
lines of constant eccentricity are plotted against $\zeta$ as dashed
lines.  A solid line shows all 3PN configurations having a constant
value of $m\Omega = 0.061355$.  This value of $m\Omega$ was chosen
because it corresponds to the minimum of one of the numerical EP
curves.  The horizontal dotted line displays the angular momentum
from the numerical data set.  As can be seen, there are two values of
$\zeta$ where these two lines intersect, and both correspond to valid
roots of the equation.  The smallest positive root
corresponds to a very small eccentricity which we would expect for the
given data set.  The second smallest positive root yields an
eccentricity somewhere between $\EJ = 0.25$ and $\EJ = 0.50$.  It is
unlikely that the minimum of an EP curve would have such high
eccentricities and negative values of $\zeta$ aren't allowed, so we
always choose the smallest positive root for $\zeta$.  The 3PN energy
from Eq.~(\ref{eq:total_E}) yields similar results as can be seen in
Fig.~\ref{fig:3PN_lines}.  Again, we always choose the smallest
positive root for $\zeta$.

\begin{figure}[th]
\includegraphics[width=\linewidth,clip]{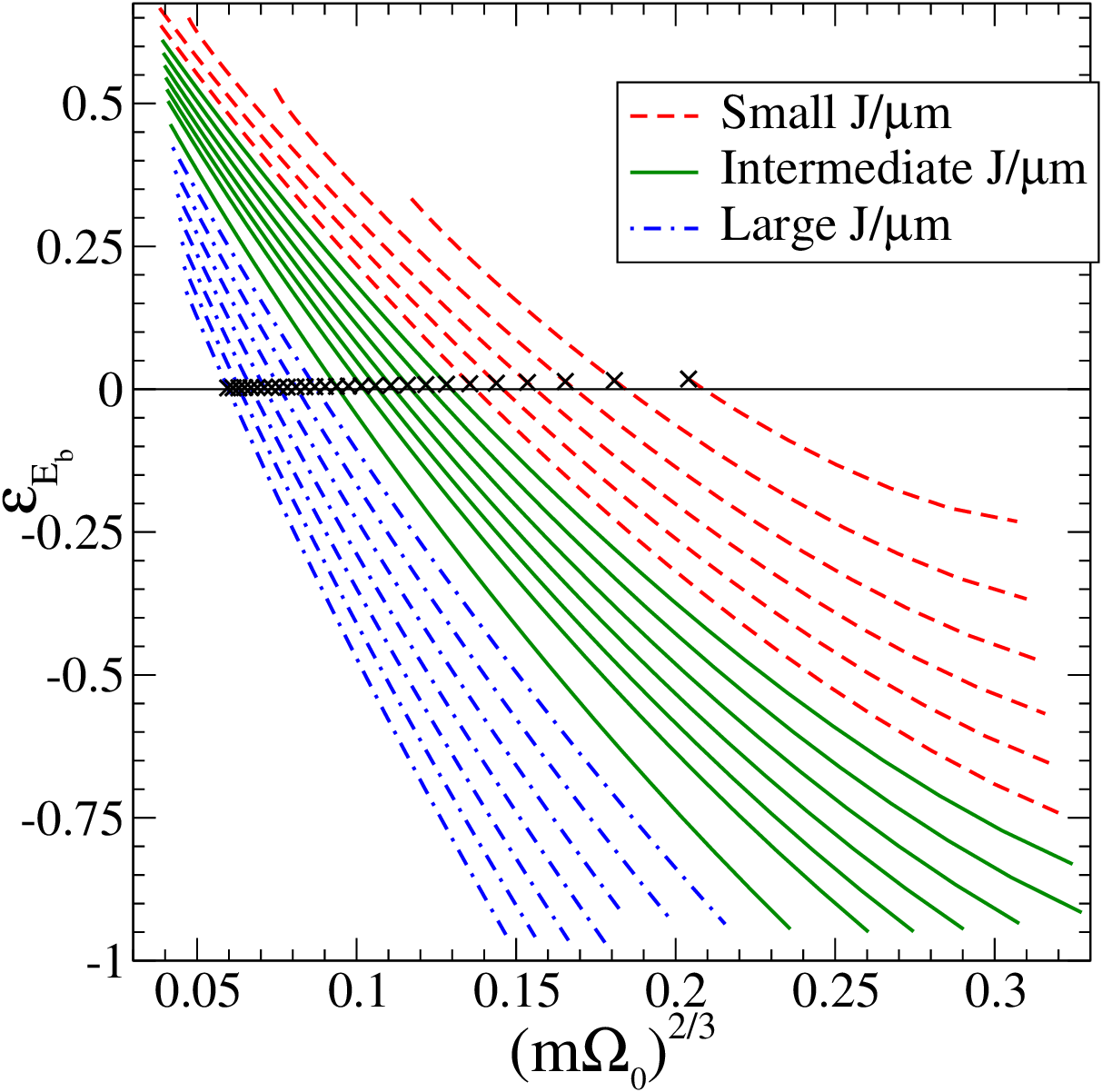}
\caption{\label{fig:Non_3PN_ConstJ} Eccentricity of non-spinning
equal-mass black holes computed along numerical EP curves.  The 3PN
$\EE$ definition of eccentricity is plotted against the orbital
angular velocity.  Lines are as in Fig.~\ref{fig:non_newt}.  The
$\times$ symbols mark the minima of each EP curve.}
\end{figure} 

In Fig.~\ref{fig:Non_3PN_ConstJ} we show the eccentricity $\EE$ of the
same EP curves considered in Fig.~\ref{fig:non_newt} but computed
using the 3PN energy equation.  The minima of the EP curves, which
would have $\EO=0$, are marked with $\times$'s. Qualitatively, these
results resemble the previous direct measurements shown if
Fig.~\ref{fig:non_newt}, however there are differences.  First, note
that the the minima of the EP curves do not correspond exactly to
$\EE=0$ (as first noticed in 
Refs.\cite{BertiIyerWill-2006a,mora-will-2002,mora-will-2004}).  
Also, for high values of $J/\mu m$, $\EE$ can be evaluated for
more of the data points on these EP curves than is possible for $\EO$.
The evaluation of $\EO$ was limited in this range because the
numerical data did not extend out to sufficiently large separations
and $\EO$ requires pairs of turning points to measure the
eccentricity.  Because $\EE$ requires information from only a single
initial-data set, it can be computed for some data points where $\EO$
cannot.  While not presented here, using the angular momentum equation
to compute $\EJ$ delivers results that are qualitatively similar.

\begin{figure}[th]
\includegraphics[width=\linewidth,clip]{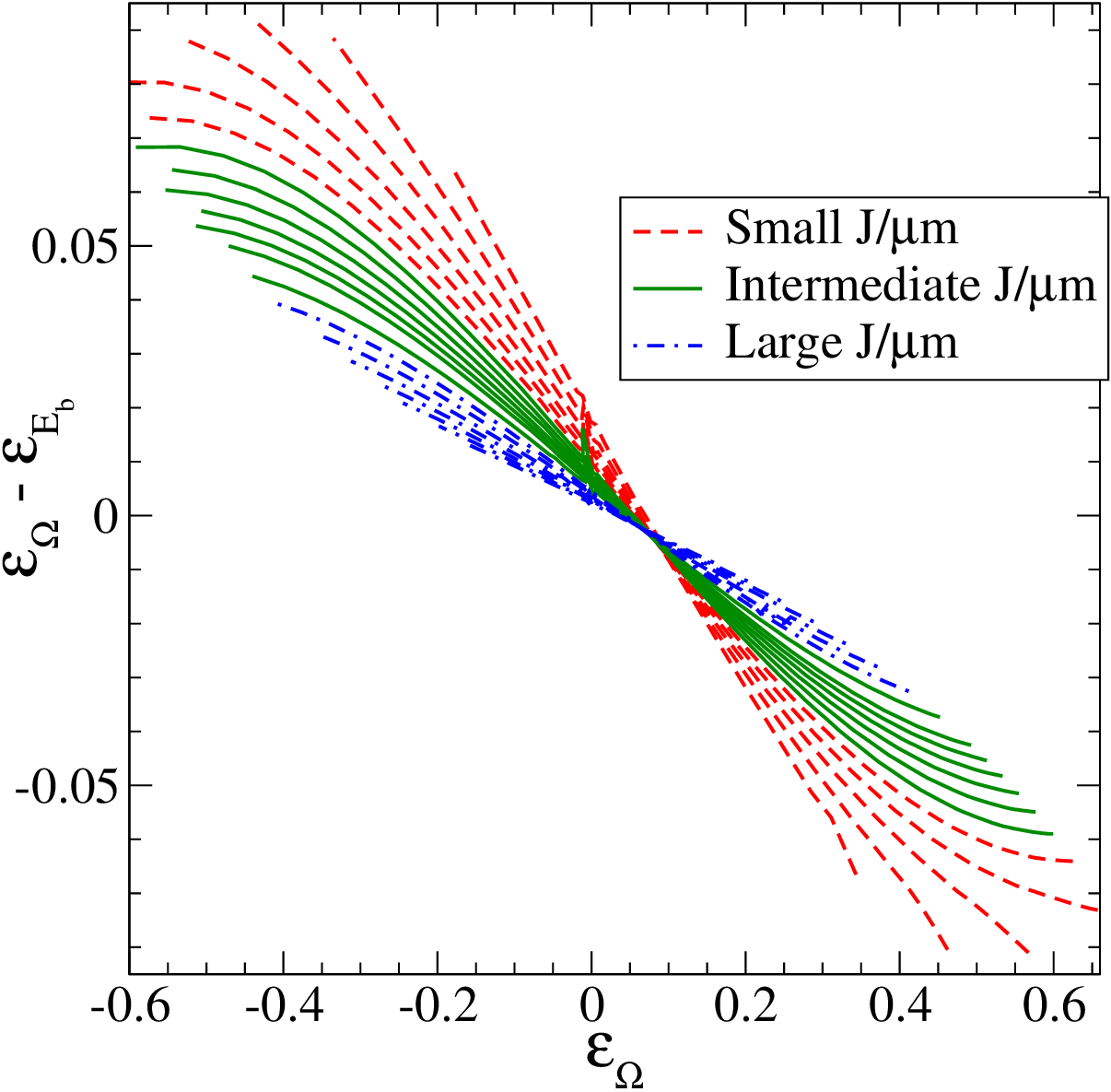}
\caption{\label{fig:Non_3PN_Om} Difference between $\EO$ and $\EE$
plotted against $\EO$ for non-spinning equal-mass black holes computed
along numerical EP curves.  Lines are as in Fig.~\ref{fig:non_newt}.}
\end{figure}

Finally, our goal has been to gauge whether or not $\EO$ was a reasonable
definition of eccentricity.  In Fig.~\ref{fig:Non_3PN_Om}, we show the
difference between the $\EO$ and 3PN definition of eccentricity
$\EE$.  We plot $\EO - \EE$ against $\EO$ rather than a relative
difference to avoid division by small number issues that make the
graph difficult to read (recall that $\EO$ and $\EE$ do not evaluate
to zero for the same data points). As expected, there is better
agreement for more Newtonian configurations (large $J/\mu m$) which
diminishes as $J/\mu m$ decreases.  There is some jaggedness at $\EO =
0.0$ caused by the polynomial fitting used to estimate the minima
of the EP curves.  A careful examination of the apocenter data points
(positive $\EO$) 
shows that relative errors for modest values of $\EO$ do not exceed 
20\% for even the smallest values of $J/\mu m$.  Of course, the relative
errors near $\EO=0$ are unbounded.  This comparison suggests that the use of
either $\EO$ or $\EE$ yield reasonable measures for the eccentricity
when applied to the numerical initial-data sets, although we should be
more cautious in trusting results for pericenter data with small values
of the angular momentum.

\subsection{The Komar-Mass Difference}

\begin{figure}[ht]
\includegraphics[width=\linewidth,clip]{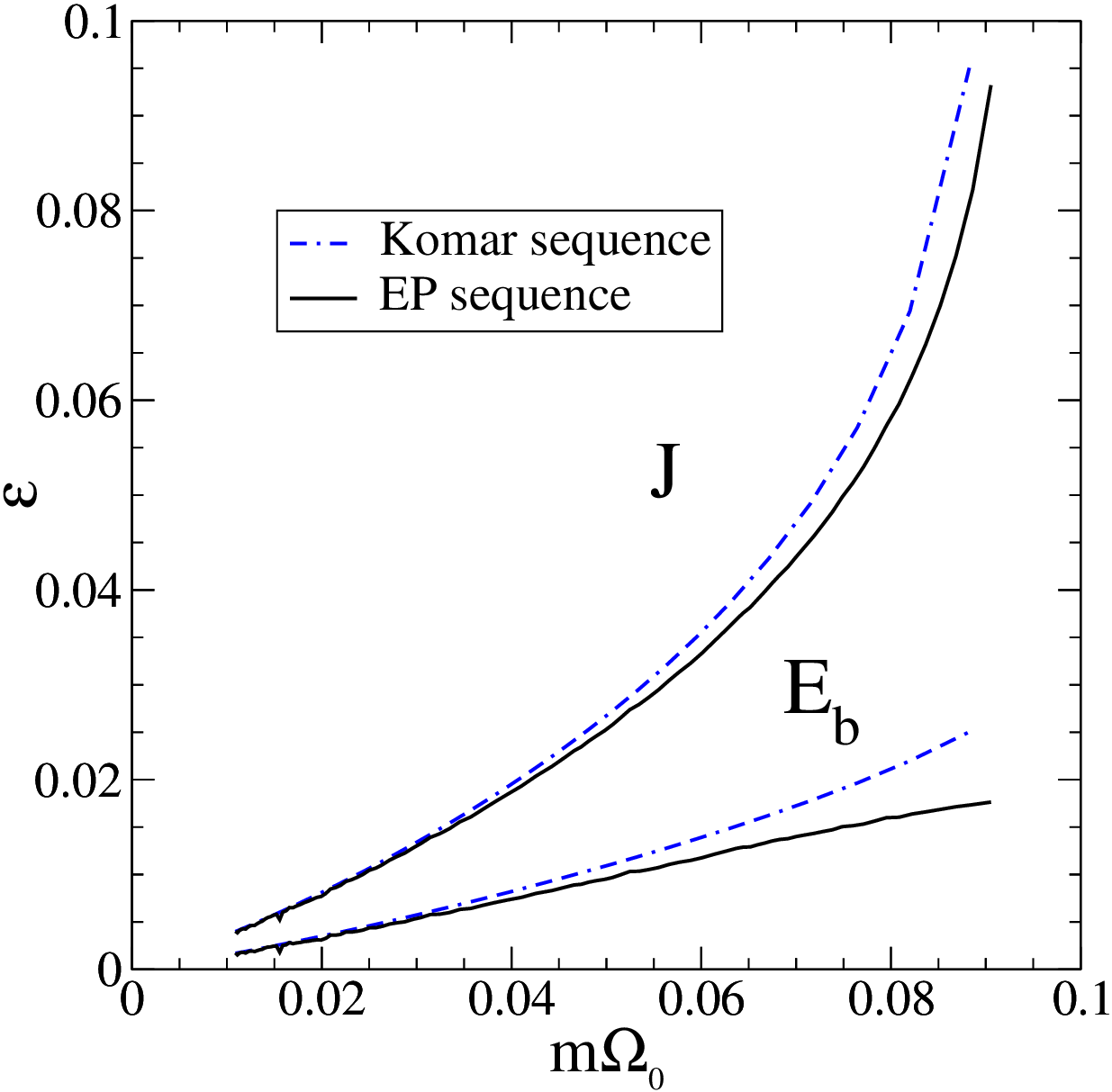}
\caption{\label{fig:3PN_Non_Komar} The 3PN eccentricity measures $\EE$
and $\EJ$ applied to both Komar and EP sequences of non-spinning
equal-mass black holes.  Note that the EP minima yield quasi-circular
data with a smaller eccentricity than is obtained from the Komar-mass
ansatz.}
\end{figure}

The first application of the $\EE$ and $\EJ$ definitions to
non-spinning, equal-mass black-hole binary initial data was undertaken
by Berti et.\ al.\cite{BertiIyerWill-2006a}.  In this work, they
considered initial data sets that satisfy the Komar-mass criteria for
circular orbits and showed that the 3PN definitions of eccentricity
$\EE$ and $\EJ$ yield non-zero results for these configurations which
are supposed to be in circular orbits.  We reproduce these results in
Fig.~\ref{fig:3PN_Non_Komar}.  We note that, while it is true that the
circular orbit data have non-vanishing 3PN eccentricity, the magnitude
of this eccentricity smoothly approaches zero as the separation
increases.  We also note, as pointed out in
Refs.~\cite{mora-will-2004} and \cite{BertiIyerWill-2006a}, that the
energy based definition $\EE$ yields consistently smaller values of
eccentricity for the ``quasi-circular'' data than does the angular
momentum based definition $\EJ$.

In Ref.~\cite{caudill-etal-2006}, we showed that circular orbits
defined by the EP method yield models that are very similar to those
defined by the Komar-mass ansatz.  We also showed that, while very
similar, all quantitative measurements of the quality of the circular
orbits showed that the EP method yields better results.  This is again
true if we compare the 3PN eccentricities computed for non-spinning,
equal-mass binaries in circular orbits defined by the EP method
and the Komar-mass ansatz.  The results are shown in
Fig.~\ref{fig:3PN_Non_Komar}, where it is clear that the the EP method
yields consistently smaller values of eccentricity for quasi-circular orbit
data.

Although the EP method yields consistently better results for circular
orbits than can be obtained using the Komar-mass method, the
differences are in general not sufficiently significant to outweigh
the considerable computational expense associated with the EP method.
When we compute eccentricities using the $\EO$ definition, the
overhead of using the EP method is even larger.  It would be useful to
find another means of estimating the eccentricity of binary initial
data.

\begin{figure}[ht]
\includegraphics[width=\linewidth,clip]{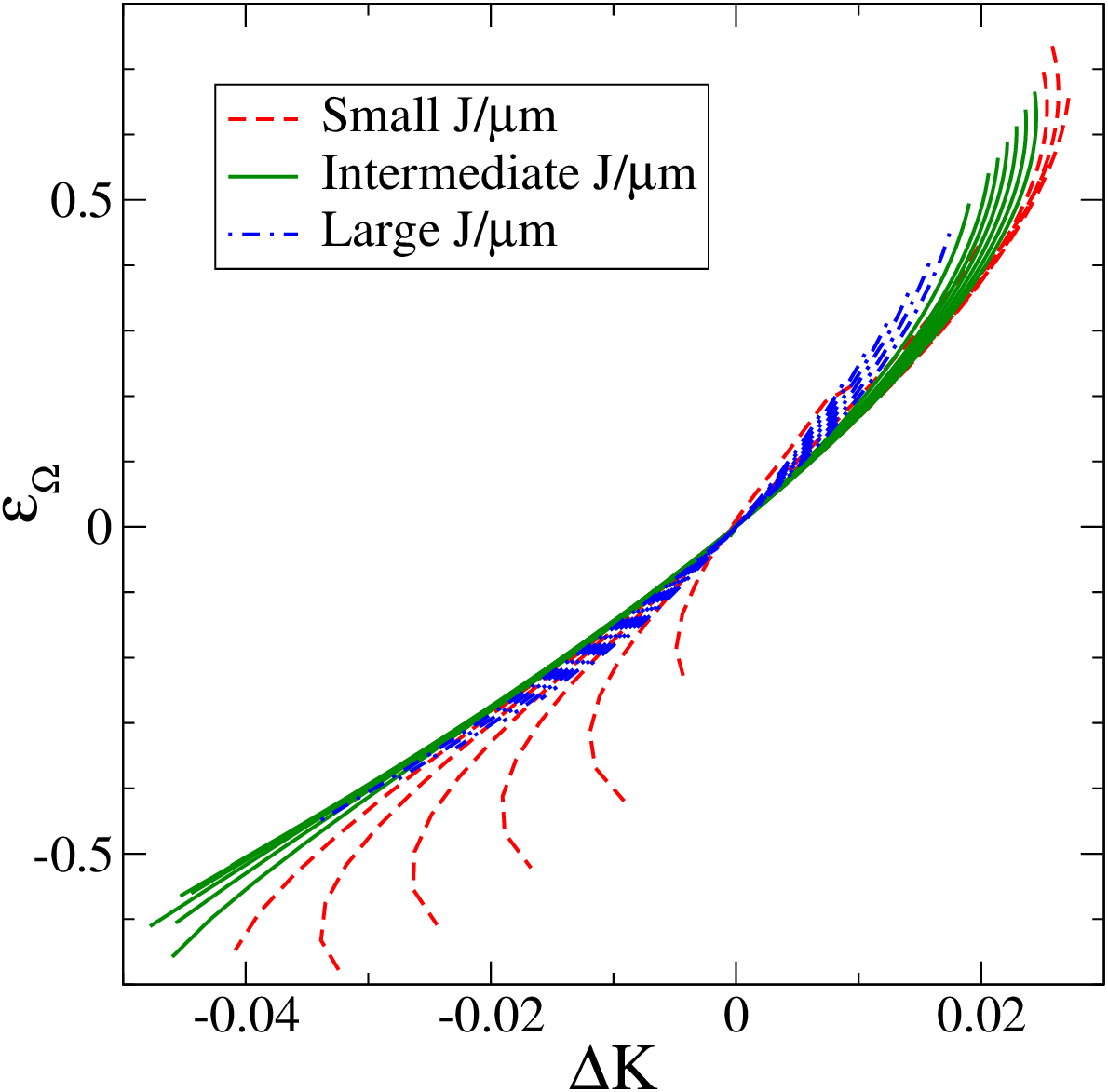}
\caption{\label{fig:non_Komar_newt} The eccentricity measure $\EO$
applied to non-spinning equal-mass black holes computed along
numerical EP curves and plotted against the Komar-mass difference
$\Delta{K}$.  Lines are as in Fig.~\ref{fig:non_newt}.}
\end{figure}

In Fig.~\ref{fig:non_Komar_newt}, we again plot the eccentricity $\EO$
of the same set of non-spinning, equal-mass initial data.  However, on
the horizontal axis, we plot the dimensionless Komar-mass difference
\begin{equation}
\Delta K \equiv \left(E_{ADM} - M_{K}\right)/\mu.
\end{equation}
We see a very strong correlation in the data, though the correlation
weakens as we move further from quasi-circular orbits. It is worth
noting that the EP curves with low $J/\mu m$ curve back towards zero
$\Delta{K}$ for configurations with negative eccentricities
(pericenter) and small $J/\mu{m}$.  This behavior is for data in the
region of the local maxima in the EP curves.  That this occurs is
consistent with the notion that these local maxima represent unstable
quasi-circular orbits, and hence we will find the Komar-mass difference
vanishing in this region.  More importantly, it shows that all of the
definitions of eccentricity we have used will break down in this
highly relativistic region.

\section{Corotation}\label{sec:corotation}

\begin{figure}[ht]
\includegraphics[width=\linewidth,clip]{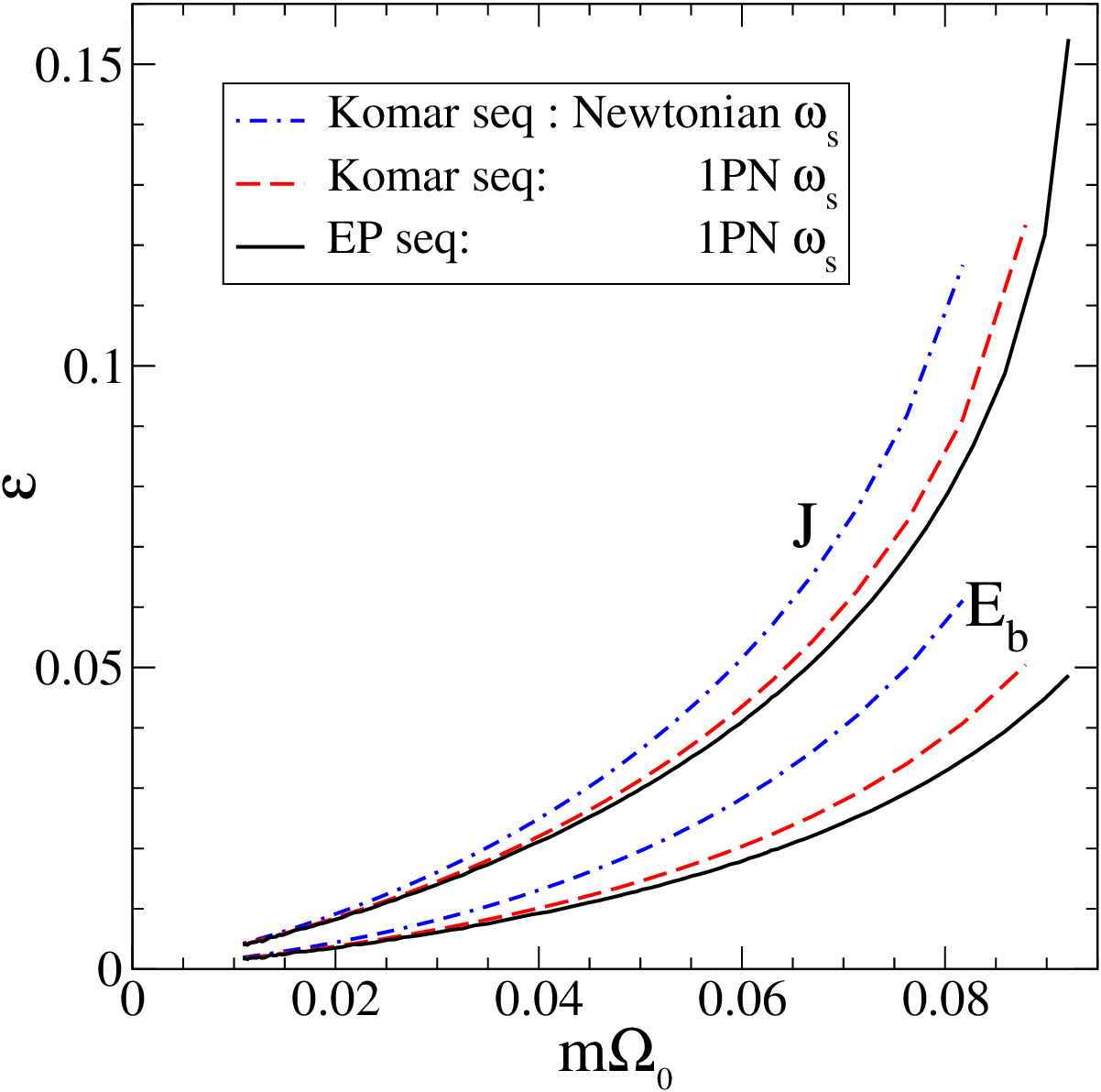}
\caption{\label{fig:Co_correction} The 3PN eccentricity measures $\EE$
and $\EJ$ applied to both Komar and EP sequences of {\em corotating}
equal-mass black holes.  The dot-dashed(blue) lines correspond to
quasi-circular data defined via the Komar-mass ansatz and where the
Newtonian notion of corotation is used in the 3PN equations.  The
dashed(red) lines show the improvement obtained by including the 1PN
correction to the notion of corotation.  The solid(black) lines show
the added improvement of using quasi-circular data based on the EP
method.}
\end{figure}

So far we have restricted ourselves to the case of non-spinning
black holes.  However, corotating configurations have received
considerable attention in spite of the fact that we do not expect to
see corotating black holes in nature.  In addition to the non-spinning
case, Berti et.\ al.\cite{BertiIyerWill-2006a} also computed the 3PN
eccentricities for the corotating data presented in
Ref.~\cite{caudill-etal-2006} and we reproduce these results in
Fig.~\ref{fig:Co_correction}.  

One of the primary reasons that the
initial data for corotating black holes has been studied so
extensively is that there is a simple and unique means of enforcing the
condition of corotation on the black holes.  This is in contrast to
any attempt to enforce a specific value of spin (even no spin) on each
black hole, which necessarily includes the uncertainty in how we define
the spin of an individual black hole in a binary configuration.

The Newtonian concept of corotation implies that each black hole
rotates with a spin angular velocity $\omega_s$ that is equal to the
orbital angular velocity.  This Newtonian notion of corotation (i.e.\
$\omega_s = \Omega_0$) was used to fix the spin parameters in
Eqs.~(\ref{eq:total_E}) and (\ref{eq:total_J}) when using these 3PN
equations to compute $\EE$ and $\EJ$.  To our knowledge, this
Newtonian notion of corotation has been used in {\em all} PN
computations dealing with corotation (cf
Refs.~\cite{BertiIyerWill-2006a,Damour-etal-2002,Blanchet:2002,blanchet2003,mora-will-2004,Campanelli_etal2006}).
However, in Ref.~\cite{caudill-etal-2006}, we have shown that there
are relativistic corrections to the spin angular velocity associated
with corotating black holes.  We find $\omega_s$, including the 1PN
correction, to be of the form
\begin{equation}\label{Eqn:spin}
\omega_s = \Omega_0 (1-\eta (m\Omega_0)^{2/3} + \cdots),
\end{equation}
where $\eta=\mu/m$ is the symmetric mass ratio which takes the value
of $1/4$ for equal-mass binaries.  Also shown in
Fig.~\ref{fig:Co_correction} are the two 3PN eccentricities computed
using the corrected definition for $\omega_s$.  We find that using
this corrected definition significantly decreases the 3PN estimated
eccentricity for these circular-orbit models.

As with the non-spinning case, we can also compute the eccentricities
for the corotating equal-mass binaries in quasi-circular orbits
defined in terms of the EP method rather than the Komar-mass ansatz.
Including also the improved definition for $\omega_s$ in the 3PN
definitions of eccentricity, we find the evaluated eccentricities are
smallest when evaluated for quasi-circular data defined by the EP
method.  This can also be seen in Fig.~\ref{fig:Co_correction}.

Even with the correction to $\omega_s$, we notice the magnitude of the
3PN eccentricities computed for ``circular'' data are consistently
larger for corotating binary data than for non-spinning data.  We
cannot be certain why this is the case, but we should keep in mind
that there is an inherent inconsistency in any attempt to attach the
notion of eccentricity to corotating configurations.  All of our
definitions for eccentricity, including the 3PN definitions that can
be evaluated using information from a single data set, ultimately rely
on information from both a pericenter and an apocenter configuration.
For eccentric orbits, the spins of corotating black holes will change
from pericenter to apocenter.  Since the spins (including spin-orbit
and spin-spin couplings) contribute to the total energy, this
variation of the spin throughout the orbit must impact upon our
definitions of eccentricity.  This is likely to cause few problems
when considering nearly circular orbits (where the spin varies little
from pericenter to apocenter), but our definitions of eccentricity may
not yield reasonable results for orbits that deviate significantly
from being circular.

\section{Conclusions}\label{sec:conclusions}

In this paper, we have examined several basic questions associated
with the construction of binary black-hole initial data.  When we
construct binary initial-data sets using the extended CTS equations
and fix the freely-specifiable parts of the data and boundary
conditions to be consistent with the assumptions of quasiequilibrium,
we obtain models for black holes in circular orbits.  But what happens
if we set aside the quasiequilibrium assumption that imposes
circular orbits?  The resulting data can no longer evolve in a
quasiequilibrium manor as the orbit will have a significant
eccentricity.  Our investigations suggest that the initial data models
we obtain represent, in general, binaries that are at {\em turning points}
(either apocenter or pericenter) of some general eccentric orbit.

The specific notion of quasiequilibrium that we set aside is
implemented by imposing either the Komar-mass condition or by
choosing the minimum of an EP curve as our circular orbit model.  If,
as it seems, general initial-data models on an EP curve are at turning
points, then we can use information from these models to estimate the
eccentricity of the model's orbits.  Of course, there is no unique
definition of eccentricity.  We have examined several possible
definitions for eccentricity.  Using only information from the
initial-data sets on an EP curve, we have defined two eccentricities
for an orbit ($\EO$ and $\EL$), but these definitions require that 
we have representative models at
both the pericenter and apocenter turning points of a given orbit
(assumed to have constant binding energy).  We have compared these
definitions of eccentricity to the 3PN definitions ($\EE$ and $\EJ$)
developed by Mora and Will\cite{mora-will-2004}.  All of the
definitions agree quite well for non-relativistic orbits.  They are
also in reasonably good agreement for more tightly-bound and
relativistic orbits as well, although the results are quantitatively
different.  

One might ask which definition is better for more relativistic
situations.  However, it isn't clear that there is a meaningful
answer.  In comparing the numerical and 3PN eccentricity measures for
relativistic cases, we have noticed an interesting and unexpected feature
of the 3PN equations.  Figure~\ref{fig:Non_const_J} compares numerical
and 3PN EP curves for equal-mass non-spinning black-hole binaries.
The EP curves each have a constant value for the orbital angular
momentum.  All of the numerical EP curves cover a finite range of
separations (parameterized by $m\Omega_0$).  This is because it
becomes computationally expensive to compute models at very large
separations (small $m\Omega_0$) and it becomes increasingly difficult
to obtain convergent solutions at very small separations.  {\em What was
unexpected is that the 3PN EP curves also cover a finite range
of separations.}

For sufficiently small values of $J/\mu{m}$, the 3PN EP curves do not
extend to configurations with arbitrarily large separation.  To make
this clear, Fig.~\ref{fig:Non_const_J} includes a curve that marks the
boundary (for both large and small separation) of the 3PN EP curves.
This is most easily seen for the 3PN EP curves near the bottom of
Fig.~\ref{fig:Non_const_J}.  Here we can see that the 3PN EP curves do
not extend to arbitrarily small values of $m\Omega_0$, but the
numerical EP curves do (although we do not compute them for
arbitrarily large separation).  

That this behavior is not an artifact of our method for computing the
3PN EP curves can be seen by examining Fig.~\ref{fig:3PN_lines_J}.
Recall that angular momentum is held constant along EP curves, so an
EP curve is represented by a horizontal line in this figure.  Notice
that all curves of constant eccentricity have a local minimum.  For
$J/\mu{m}\gtrsim3.4$, an EP curve can extend from $\epsilon=0\to1$.
However, for $J/\mu{m}\lesssim3.4$, can only extend from $\epsilon=0$
to a maximum eccentricity that is less than $1$, and which is determined
by which curve of constant $\epsilon$ has its minimum tangent to
the EP curve.

It is physically reasonable that large eccentricity orbits should not
exist as the orbits become sufficiently relativistic since the finite
size of the black holes would lead to a collision.  However, the PN
equations treat the black holes as point particles and should not be
sensitive to this limitation.  Also, we remind the reader that the
numerical EP curves have no difficulty in extending to large
separation ($\epsilon\to1$) for relativistic orbits.  This suggests
that, for relativistic orbits, there may be a problem with the
definition of eccentricity and semilatus rectum used by Mora and
Will\cite{mora-will-2004} to parameterize the PN energy and angular
momentum.  We do not consider this to be a serious problem, but rather an
indication that any definition of eccentricity is of limited value for
orbits where the interaction between the black holes becomes highly
relativistic.  

Given the strong correlation between the various
measures of eccentricity and the difference between the Komar mass and
the ADM energy as measured by $\Delta{K}$ and seen in
Fig.~\ref{fig:non_Komar_newt}, we suggest that perhaps $\Delta{K}$
can serve as a useful invariant means of parameterizing the eccentricity of
an orbit.  However, initial-data studies can at best suggest possible
useful parameterizations.  It will be most useful to evolve initial
data that are significantly eccentric and examine the orbital dynamics
to better understand both the parameterization of eccentricity and its
effects on the dynamics.  For example, it may be particularly
interesting to explore the evolution of eccentric initial data for an
``orbit'' that has no pericenter turning point on its EP curve.

\acknowledgments 

We would like to thank Harald Pfeiffer, Alessandra Buonanno, Clifford
Will, and Emanuel Berti for useful discussions.  This work was
supported by NSF Grant No. PHY-0555617 to Wake Forest University.
G.C.\ acknowledges support from the Z. Smith Reynolds Foundation.
Computations were performed on the Wake Forest University DEAC Cluster
with supported from an IBM SUR grant and the Wake Forest University IS
Department.

\end{document}